\documentclass[twocolumn]{aastex631}
\usepackage{newtxmath}

\usepackage{graphicx}
\usepackage{booktabs}
\usepackage{amsmath,amssymb}
\usepackage{hyperref}

\newcommand{\Teff}{\ensuremath{T_{\rm eff}}}
\newcommand{\logg}{\ensuremath{\log g}}
\newcommand{\feh}{\ensuremath{[{\rm Fe}/{\rm H}]}}
\newcommand{\sigdPi}{\ensuremath{\sigma_{\Delta\Pi_1}}}

\newcommand{\JKS}{\ensuremath{(J-K_s)_0}}
\newcommand{\MKs}{\ensuremath{M_{K_s}}}

\renewcommand{\textcolor}[2]{#2}
\begin{document}

\title{An All-Sky Catalog of 6.5 Million Primary Red Clump Stars\\from \textit{Gaia} DR3 XP Spectra}

\author[0000-0002-2264-745X]{Zheng Yu}
\affiliation{Institute of Astrophysics, Chuxiong Normal College, Chuxiong 675000, P.~R.~China}
\email{yuzheng\_hs@163.com}

\author[0000-0003-2472-4903]{Bingqiu Chen}
\affiliation{South-Western Institute for Astronomy Research, Yunnan University, Kunming 650500, P.~R.~China}
\email{bchen@ynu.edu.cn}

\correspondingauthor{Zheng Yu, Bingqiu Chen}

\begin{abstract}
Red clump (RC) stars are excellent standard candles for mapping the three-dimensional structure of the Milky Way. We construct an all-sky catalog of 6.5 million primary RC stars \textcolor{blue}{using spectroscopic estimates of asteroseismic parameters inferred from \textit{Gaia} DR3 low-resolution XP spectra. We train a mixture density network (MDN) on a cross-matched sample. The network maps each 343-dimensional corrected XP spectrum to estimates of $\Delta\Pi_1$, the asymptotic period spacing of dipole gravity modes, and $\Delta\nu$, the large frequency separation. We select primary RC stars using these two estimates.} We release two complementary catalogs defined by different parameter-space cuts and thresholds: a high-purity Tier~1 sample of $532{,}189$ stars with 97\% purity and a high-completeness Tier~2 sample of $6{,}534{,}931$ stars with 86\% completeness. Both catalogs cover the full celestial sphere. We provide distances and extinction values for all stars. We obtain a median photometric distance precision of 6\% for Tier~1 RC stars. Using the resultant distances, we independently calibrate the \textit{Gaia} DR3 parallax zero-point offset. The three-dimensional density distribution traced by the Tier~2 sample extends continuously to $R \simeq 25$\,kpc. Both catalogs are publicly available. These catalogs provide a valuable resource for Galactic archaeology, delivering a homogeneous dataset to trace the chemo-dynamical evolution of the Milky Way and to calibrate models of stellar populations.
\end{abstract}


\section{Introduction}
\label{sec:intro}

Mapping the three-dimensional structure, chemical abundance patterns, and kinematic properties of the Milky Way is fundamental to understanding its formation and evolution \citep{bland2016galaxy}. Such studies require large stellar samples with precise distance measurements that can probe from the inner disk to the distant outskirts. Red clump (RC) stars, with their tightly clustered absolute magnitudes, serve as premier standard candles \citep{salaris2002population,girardi2016red,huang2020mapping,yu2025empirically}.

Traditional approaches select RC stars as an over-density in color--magnitude diagrams \citep{stanek1997modeling,lopez2002old}. The primary challenge in constructing a pure RC sample is contamination from red giant branch (RGB) stars. These stars overlap with the RC in both color and absolute magnitude, leaving substantial residual contamination. Large-scale spectroscopic surveys, such as the Large Sky Area Multi-Object Fiber Spectroscopic Telescope (LAMOST; \citealt{cui2012large,zhao2012lamost}) and the Apache Point Observatory Galactic Evolution Experiment (APOGEE; \citealt{majewski2017apache}), have enabled RC selection in the \Teff--\logg\ plane. \citet{bovy2014apogee} defined \Teff--\logg\ boundaries calibrated against \textit{Kepler} asteroseismology to select RC stars from APOGEE data. \citet{huang2020mapping} introduced a similar metallicity-dependent boundary, yielding $\sim$140,000 primary RC stars from LAMOST. \citet{wu2026mapping} extended this to $\sim$8.4 million RC candidates using atmospheric parameters derived from \textit{Gaia} BP/RP spectra \citep{andrae2023b}. The purity and completeness of the derived primary RC sample depend on the accuracy of the adopted stellar atmospheric parameters. For parameters from LAMOST spectra, these values are approximately 80\%, whereas they decrease when atmospheric parameters are derived from \textit{Gaia} XP data \citep{huang2020mapping,wu2026mapping}.

\textcolor{blue}{Asteroseismology provides observables that can separate red-clump (RC) stars from red-giant-branch (RGB) stars \citep{chaplin2013asteroseismology}. Two such quantities are the observed period spacing $\Delta P$ and the asymptotic gravity-mode period spacing $\Delta\Pi_1$. $\Delta P$ is measured directly between consecutive visible dipole mixed modes in the oscillation spectrum. Because these mixed modes behave as pressure modes in the envelope and as gravity modes in the core, their local period spacing varies across an acoustic radial order and depends on the mode coupling and visibility. This makes $\Delta P$ a useful but empirical separator of RC and RGB stars \citep{beck2011kepler,bedding2011gravity,mosser2011mixed}. However, the observed $\Delta P$ is systematically smaller than the underlying asymptotic dipole gravity-mode spacing $\Delta\Pi_1$ \citep{beck2011kepler,bedding2011gravity,mosser2011mixed,mosser2012probing}. By contrast, $\Delta\Pi_1$ is the asymptotic spacing of pure dipole gravity modes, which is a more fundamental property of the stellar core. It is inferred by fitting the mixed-mode asymptotic relation or by stretching the oscillation spectrum so that gravity-dominated modes become nearly evenly spaced \citep{mosser2012probing,vrard2016period}. The inferred $\Delta\Pi_1$ also cleanly separates RC from RGB stars, and in practice it provides a more robust classification than the directly measured $\Delta P$, especially for stars near the boundary between the two evolutionary states.}

\textcolor{blue}{The large frequency separation $\Delta\nu$ primarily traces mean stellar density. Inferring stellar mass from global seismic scaling relations also requires the frequency of maximum oscillation power $\nu_{\max}$ and the effective temperature. After the core-helium-burning branch has been isolated using $\Delta\Pi_1$, higher-mass secondary-clump stars tend to occupy higher $\Delta\nu$ at similar $\Delta\Pi_1$. We therefore use $\Delta\nu$ only as an empirical screening quantity, not as a direct mass measurement or a universal physical boundary \citep{Kjeldsen1995,mosser2014mixed,pinsonneault2018second}.}

\textcolor{blue}{The APOGEE--\textit{Kepler} Asteroseismology Science Consortium (APOKASC) program provides global seismic parameters and evolutionary classifications for thousands of red giants \citep{pinsonneault2014apokasc,pinsonneault2018second}. \citet{vrard2016period} published asymptotic $\Delta\Pi_1$ measurements for more than 6,100 \textit{Kepler} red giants. \citet{yu2018asteroseismology} published homogeneous $\nu_{\max}$ and $\Delta\nu$ measurements for 16,094 red giants, together with derived stellar properties. The target list of \citet{yu2018asteroseismology} combined six previously published samples \citep{hekker2011characterization,huber2011testing,stello2013asteroseismic,huber2014revised,mathur2016probing,yu2016asteroseismology}, whereas the recommended evolutionary-state flags were compiled from several independent seismic classifications \citep{bedding2011gravity,stello2013asteroseismic,mosser2014mixed,vrard2016period,elsworth2017new,hon2017deep}. More recently, \citet{vrard2025status} consolidated several independent seismic classification methods for the full \textit{Kepler} red-giant sample, explicitly retaining the distinction between methods based on $\Delta P$ and those based on $\Delta\Pi_1$. However, these direct seismic measurements and classifications remain confined to the narrow sky footprints of space-based time-domain missions.}

Stellar spectra can be used to infer \textcolor{blue}{$\Delta\Pi_1$} and $\Delta\nu$, because the surface carbon-to-nitrogen ratio [C/N]---altered by the first dredge-up \citep{iben1967stellar,salaris2005evolution} in a mass-dependent manner---imprints itself on molecular absorption features \citep{salaris2015post,martig2016red}. \citet{hawkins2018photospheric} demonstrated that CN bands in APOGEE spectra could be used to infer seismic parameters. \textcolor{blue}{\citet{ting2018large} developed a neural-network framework that transferred asteroseismic labels to a large sample of giants from APOGEE and LAMOST. After restricting the LAMOST portion to ${\rm S/N}_{\rm pix} > 75$, their duplicate-corrected combined sample contains 85,539 primary RC stars with $\sim$3\% contamination \citep{ting2018erratum}, superseding the 92,249-star figure quoted in the original Letter.} \citet{wang2023precise} extended this to LAMOST DR8, classifying 180,436 primary RCs with purities $>95\%$. \citet{lucey2020inner} showed that this approach works even with broadband photometry, producing $\sim$2.6 million RC candidates from 2MASS, AllWISE, Pan-STARRS, and \textit{Gaia} photometry, albeit with higher contamination.

In this work, we perform an all-sky RC selection using \textit{Gaia} DR3 BP/RP (XP) spectra \citep{collaboration2023gaia,de2023gaia,montegriffo2023gaia}. Despite their low resolution ($R \simeq 30$--100), XP spectra capture the broad molecular features that carry the [C/N] signature across the full celestial sphere for $\sim$220 million stars. \textcolor{blue}{We train a mixture density network (MDN; \citealt{bishop1994mixture}) on the cross-matched catalog of \citet{ting2018large} to infer spectroscopic estimates of $\Delta\Pi_1$ and $\Delta\nu$ from the 343-dimensional corrected XP spectra of \citet{huang2024comprehensive}. The MDN provides a selected-component point estimate and an associated within-component width, which we use as an empirical quality-ranking statistic for sample selection.} Applying different seismic evolutionary cuts and \textcolor{blue}{quality thresholds}, we construct two complementary all-sky primary RC catalogs: a high-purity Tier~1 and a high-completeness Tier~2.

\section{Data}
\label{sec:data}

This work rests on three core data products. The all-sky \textit{Gaia} DR3 XP spectra serve as the direct input features for the MDN. The \citet{zhang2025three} atmospheric parameter catalog provides \Teff, \logg, and \feh\ for approximately $2.2 \times 10^8$ stars, forming the parametric basis for defining the giant-star parent sample. \textcolor{blue}{The \citet{ting2018large} catalog supplies the $\Delta\Pi_1$ and $\Delta\nu$ labels used to train the mapping from XP spectra to asteroseismic parameters.}

The post-sampled XP continuous spectra from \textit{Gaia} DR3 \citep{collaboration2023gaia} provide flux-calibrated low-resolution spectra covering 336--1020\,nm at $R \simeq 30$--100 for approximately 220 million stars \citep{de2023gaia,montegriffo2023gaia}. \textit{Gaia} DR3 itself delivers stellar parameter estimates via the General Stellar Parametrizer from Photometry (GSP-Phot) module \citep{andrae2023a}, but their accuracy is limited by systematic errors in the forward-modeling of low-resolution spectra. The raw XP spectra exhibit color- and magnitude-dependent systematic errors at the BP/RP junction and in the near-ultraviolet \citep{montegriffo2023gaia,huang2024comprehensive}. To eliminate the impact of these systematics on downstream neural-network feature extraction, we use the 343-pixel corrected XP continuous spectra of \citet{huang2024comprehensive}. Their flux correction is anchored to CALSPEC \citep{bohlin2014techniques,bohlin2022update} and the Next Generation Spectral Library (NGSL; \citealt{heap2007hubble}) as absolute-flux references, with LAMOST DR7 (\citealt{luo2015first}) used to extend the $G$-magnitude term of the correction to the faint end ($G \lesssim 17.5$). The correction was validated against two independent empirical libraries, the Medium-resolution Isaac Newton Telescope Library of Empirical Spectra (MILES; \citealt{sanchez2006medium}) and the Library of Empirical Medium-resolution spectra by Observations with the NAOC Xinglong 2.16~m and YNAO Gaomeigu 2.4~m telescopes (LEMONY; \citealt{wang2018lemony}), showing a consistency better than $\sim 2\%$ over 336--400~nm and $\sim 1\%$ at redder wavelengths.

To define a clean giant-star parent sample for downstream inference, we adopt the all-sky atmospheric parameter catalog of \citet{zhang2025three}. Building on the forward-modeling framework of \citet{zhang2023parameters}, \citet{zhang2025three} introduce a star-by-star variable extinction curve and simultaneously determine the stellar parameters and extinction curves by jointly modeling \textit{Gaia} XP spectra with 2MASS $J/H/K_s$ and unWISE $W1/W2$ photometry under a Bayesian framework. LAMOST DR8 (\citealt{cui2012large}; \citealt{zhao2012lamost};
\citealt{luo2015first}) and Hot Payne (\citealt{xiang2022stellar}) parameters serve as priors. The resulting catalog delivers \Teff, \logg, \feh, corrected parallaxes, extinction $E$, and extinction-curve shape parameter $\xi$ for approximately 220 million stars. In the giant-parameter domain relevant to this work, \citet{zhang2025three} validated their \Teff\ and \logg\ against a held-out LAMOST validation set unseen during training, and their \feh\ against APOGEE high-resolution spectroscopy ($R \approx 22{,}500$; \citealt{majewski2017apache}), finding good agreement in both cases. When building the giant-candidate parent sample we require all sources to satisfy \textit{Gaia} DR3 renormalized unit weight error ${\rm RUWE} < 1.4$, to reject unresolved binaries whose astrometric solutions significantly deviate from the single-star model \citep{lindegren2021gaia1}. 

To establish a supervised mapping from XP continuous spectra to asteroseismic parameters, we use the catalog of \citet{ting2018large} as reference labels. \citet{ting2018large} built this catalog by transferring the large frequency separation $\Delta\nu$ and the period spacing \textcolor{blue}{$\Delta\Pi_1$} of \citet{vrard2016period} onto APOGEE DR14 and LAMOST DR3 spectra, adopting the spectroscopic atmospheric parameters ($T_{\rm eff}$, $\log g$, [Fe/H]) from those same surveys. \textcolor{blue}{The transferred period-spacing label is the asymptotic dipole gravity-mode spacing $\Delta\Pi_1$ reported by \citet{vrard2016period}, although \citet{ting2018large} denote it as ``$\Delta P$''; we correct that notation throughout the present work. We adopt the erratum-corrected version of this catalog \citep{ting2018erratum}. In this version, duplicate LAMOST visits of the same star are removed by retaining only the observation with the highest signal-to-noise ratio. This procedure yields 347,727 LAMOST stars with inferred asteroseismic parameters. The associated APOGEE catalog is unaffected because duplicate observations had already been accounted for in the original Letter.} In this work, we adopt two complementary asteroseismic quantities. \textcolor{blue}{$\Delta\Pi_1$} directly probes the core density gradient, distinguishing core-helium-burning RC stars from RGB stars with degenerate helium cores and hydrogen shell burning. \textcolor{blue}{Within the core-helium-burning population, we use $\Delta\nu$ as an empirical screening quantity to separate lower-mass primary RC stars from higher-mass secondary-clump stars.}

\section{Method}
\label{sec:method}

Our method consists of two parts. Part one (\S\ref{sec:rc_id}) focuses on physically identifying primary RC stars from the \citet{zhang2025three} all-sky giant parent sample and part two (\S\ref{sec:dist_method}) assigns distances to the identified RC sample.

\subsection{Identification of Primary Red Clump Stars}
\label{sec:rc_id}

\subsubsection{Mixture Density Network Training}
\label{sec:mdn}

\textcolor{blue}{To establish a supervised mapping from XP continuous spectra to asteroseismic parameters, we begin with the 175,512-star APOGEE--LAMOST reference-label sample selected from the erratum-corrected \citet{ting2018large} catalog \citep{ting2018erratum} after applying the adopted label-quality criteria and spectral signal-to-noise requirement. This sample contains stars from the relevant giant evolutionary stages and is distinct from the primary-RC subset quoted in the Introduction. We then retain only stars with successful cross-matches to both the \citet{huang2024comprehensive} corrected XP spectra and the \citet{zhang2025three} catalog. These cross-matches yield the final supervised modeling sample of 124,513 stars with homogeneous XP spectra and ($\Delta\Pi_1$, $\Delta\nu$) labels. No pre-erratum duplicate LAMOST observations enter this sample. For each star, the 343-dimensional XP continuous-spectrum flux vector is the input feature, and ($\Delta\Pi_1$, $\Delta\nu$) is the supervised target.}

With the training sample prepared, we construct a MDN to simultaneously infer asteroseismic parameters from XP spectra. The \textcolor{blue}{$\Delta\Pi_1$} values of RC and RGB stars form a bimodal conditional distribution. A scalar regression network minimizing mean squared error in this bimodal regime collapses to the conditional mean between the two peaks, predicting physically nonexistent intermediate states. To avoid this degeneracy, we adopt the MDN proposed by \citet{bishop1994mixture} and previously introduced to the asteroseismic--spectral inference task by \citet{lucey2020inner}. The MDN directly models the conditional probability density $p(y \mid \mathbf{x})$ of the target parameter given the input. \citet{lucey2020inner} used multi-band photometric spectral energy distributions (SEDs) and parallaxes as input; we replace the input with the 343-dimensional \textit{Gaia} XP continuous spectrum. 

Unlike standard regression networks that output a deterministic scalar point estimate, the MDN models the output as a weighted mixture of Gaussian components, with the probability density function:
\begin{equation}
p(y \mid \mathbf{x}) = \sum_{k=1}^{K} \pi_k(\mathbf{x})\, \mathcal{N}\bigl(y \mid \mu_k(\mathbf{x}), \sigma_k^2(\mathbf{x})\bigr),
\label{eq:mdn}
\end{equation}
where $\pi_k(\mathbf{x})$, $\mu_k(\mathbf{x})$, and $\sigma_k(\mathbf{x})$ are the mixing weight, mean, and standard deviation of the $k$-th Gaussian component, respectively, all output by the neural network given the input spectrum $\mathbf{x}$. For \textcolor{blue}{$\Delta\Pi_1$} we use $K = 2$ Gaussian components to explicitly model the bimodal distribution of RGB (\textcolor{blue}{$\Delta\Pi_1$} $\lesssim 100$ s) and RC (\textcolor{blue}{$\Delta\Pi_1$} $\gtrsim 250$ s) stars; for $\Delta\nu$ we adopt $K = 1$, a single-peaked Gaussian.

The network backbone is a fully connected feed-forward network: the 343-dimensional XP flux, standardized to zero mean and unit variance, is passed through three hidden layers of 256, 128, and 64 units with ReLU activations, with dropout ($p = 0.2$) applied after the first two layers. The shared 64-dimensional representation then bifurcates into independent \textcolor{blue}{$\Delta\Pi_1$} and $\Delta\nu$ output heads. The \textcolor{blue}{$\Delta\Pi_1$} head outputs the mixing weights $\pi_k$ (via softmax), means $\mu_k$, and standard deviations $\sigma_k$(via a softplus transform) of its $K = 2$ Gaussian components, while the $\Delta\nu$ head outputs the mean and standard deviation of its single Gaussian. The total number of parameters is approximately $1.3 \times 10^{5}$.

The network is trained by minimizing the negative log-likelihood (NLL):
\begin{equation}
\mathcal{L} = -\frac{1}{N}\sum_{i=1}^{N} \ln\left[ \sum_{k=1}^{K} \pi_k(\mathbf{x}_i)\, \mathcal{N}\bigl(y_i \mid \mu_k(\mathbf{x}_i), \sigma_k^2(\mathbf{x}_i)\bigr) \right].
\label{eq:nll}
\end{equation}
The joint loss for \textcolor{blue}{$\Delta\Pi_1$} and $\Delta\nu$ is the sum of the two-channel NLL. We train with the Adam optimizer \citep{kingma2014adam}, an initial learning rate of $10^{-3}$, and batch size 256, for up to 150 epochs, applying gradient-norm clipping at each step to prevent exploding gradients. \textcolor{blue}{We randomly partition the 124,513-star supervised modeling sample into training, validation, and independent blind-test subsets at an 80:10:10 ratio using a fixed random seed of 42. We apply early stopping based on the validation-set NLL, with a patience of 15 epochs.} To prevent overfitting, we apply dropout ($p = 0.2$).

\subsubsection{Application to the \citet{zhang2025three} Catalog}
\label{sec:apply_zhang}

After the MDN is trained and validated on the training set, we deploy it to the \citet{zhang2025three} catalog for large-scale inference. When applying the trained MDN, samples outside the training domain, such as dwarfs and hot stars, must be filtered to prevent the network from producing meaningless extrapolations, a known pitfall of data-driven stellar-spectroscopic methods \citep{ness2015cannon,ting2018large}. The \citet{zhang2025three} catalog contains approximately 220 million stars, and not all are suitable as RC candidates for asteroseismic inference. We therefore adopt a multi-step filtering strategy, restricting MDN application to candidate sources within the giant-parameter domain.

Firstly, we take the extrema of the training set in the three dimensions of \Teff, \logg, and \feh, and expand them outward by buffer margins of 150\,K, 0.25\,dex, and 0.5\,dex, respectively, to accommodate edge populations. This orthogonal bounding box (Box-mask) efficiently removes the dominant non-giant background from the whole sample, retaining approximately 43 million stars for the next stage of fine filtering. Secondly, we construct a 3D convex hull of the training sample in the standardized (\Teff, \logg, \feh) space. We retain only the boundary vertices of the convex hull ($\approx 100$ points), and evaluate each candidate source with \texttt{find\_simplex} to determine whether it lies inside the convex hull; only those judged to be interior are retained. This yields $\sim$22 million giant candidates that enter the MDN inference pipeline.

\textcolor{blue}{For each input spectrum, we select the Gaussian component of the $\Delta\Pi_1$ distribution with the largest mixing weight. We then report its peak location as $\Delta\Pi_{1,\rm pred}$ and its standard deviation as \sigdPi. We do not summarize the full bimodal mixture with a single global estimate, because such a summary could fall between the RGB and RC branches and therefore be physically ambiguous. The selected component is Gaussian, so its mean and median are identical, and its one-standard-deviation range corresponds to the nominal central 68\% probability interval, conditional on that component. This interval describes the spread of the adopted component rather than an empirically calibrated coverage. The mixing weight reflects the ambiguity between the two evolutionary branches, whereas $\sigma_{\Delta\Pi_1}$ measures the width of the selected conditional Gaussian component and is used as a per-star quality-ranking statistic. In contrast, the $\Delta\nu$ output contains only one Gaussian component, so no component selection is needed.}

\subsection{Distance and Extinction Determination}
\label{sec:dist_method}

The preceding steps complete the physical identification of primary RC stars. In order to use this identified sample for studies of Milky Way structure, precise distances are required. We therefore derive their distances and extinction values using a method similar to that of \citet{yu2025stellar}.

To perform a data-driven empirical calibration of the $K_s$-band absolute magnitude $\MKs$ and intrinsic color index $\JKS$ of primary RC stars, we extract a local high-fidelity calibrator subset from the high-purity primary RC sample (i.e., Tier~1 as defined in \S\ref{sec:rc_samples}), requiring: \textit{Gaia} DR3 parallax $\varpi > 1.0$\,mas with relative parallax error $\sigma_\varpi/\varpi < 0.02$, Galactic latitude $|b| \geq 30^\circ$, reddening $E(B-V) < 0.05$ from the dust map of \citet[][hereafter SFD]{schlegel1998maps}, and Two Micron All Sky Survey (2MASS; \citealt{skrutskie2006two}) $K_s$ photometric error $\sigma_{K_s} < 0.03$\,mag. The final calibrator set contains 4,820 local stars for calibration.

For each calibrator, we compute the intrinsic color $\JKS$ and the $K_s$-band absolute magnitude. We adopt the inverse of the \textit{Gaia} DR3 parallax as the distance and derive the extinction from the SFD reddening map, applying the near-infrared extinction coefficients of \citet{yuan2013empirical}. We then model the dependence of these quantities on the atmospheric parameters (\Teff, \logg, \feh) from \citet{zhang2025three} using polynomial fits. To reduce the influence of outliers, we perform the fit through a data-driven median-binning approach \citep{yu2025stellar}. For $\MKs$, starting from the 4,820 calibrators satisfying the above criteria, we divide the sample into 15 metallicity bins, apply $3\sigma$ clipping within each bin requiring a minimum of 5 surviving stars, and take the median and standard deviation as the median point and weight. A third-order polynomial in \feh\ is then fitted:
\begin{equation}
\MKs = -1.569 + 0.130\,\feh + 0.170\,\feh^2 - 0.180\,\feh^3,
\label{eq:mk}
\end{equation}
with an intrinsic dispersion $\sigma_{\MKs} = 0.11$~mag. This dispersion is comparable to the $\sim$0.09--0.17~mag reported for the RC in the near-infrared by previous calibrations \citep{chan2020gaia,hawkins2017red}.

For the intrinsic color $\JKS$, the calibrator criteria additionally require 2MASS $J$-band photometric error $\sigma_J < 0.03$~mag to ensure color-index reliability, reducing the calibrator count slightly to 4,412 stars. The functional form follows \citet{yu2025stellar}, with the coefficients refitted to our calibrators:
\begin{equation}
\begin{split}
\JKS &= 79.826 - 40.468\,\log \Teff + 5.148\,(\log \Teff)^2 \\
&\quad + 0.002\,\logg + 0.019\,\feh - 0.031\,\feh^2,
\end{split}
\label{eq:jk}
\end{equation}
with a multi-dimensional $3\sigma$ robust iterative fit (converging at round 4, rejecting 24 outliers) and an intrinsic color dispersion $\sigma_{(J-K_s)_0} = 0.023$~mag. The residuals show no systematic trends in \Teff, \logg, or \feh.

We adopt the reddening-correction framework of \citet{yu2025stellar}. The color excess is obtained from the observed $E(J-K_s)$ and the fitted intrinsic color $\JKS$, and the $K_s$-band extinction is given by
\begin{equation}
A_{K_s} = 0.739\, E(J-K_s),
\label{eq:ak}
\end{equation}
where the near-infrared extinction coefficients are taken from \citet{yuan2013empirical}. Distances are computed from the standard relation $d\,[\mathrm{kpc}] = 10^{\,0.2(\mu+5)}/1000$, where $\mu = K_s - \MKs - A_{K_s}$. We derive the per-star distance error by analytically expanding the distance modulus as a function of mutually independent input errors. This expansion fully accounts for the covariance between $E(J-K_s)$ and $\MKs$, $\JKS$, and the observed $J$ and $K_s$ magnitudes. 

\section{Results}
\label{sec:results}

\begin{figure*}
\centering
\includegraphics[width=\textwidth]{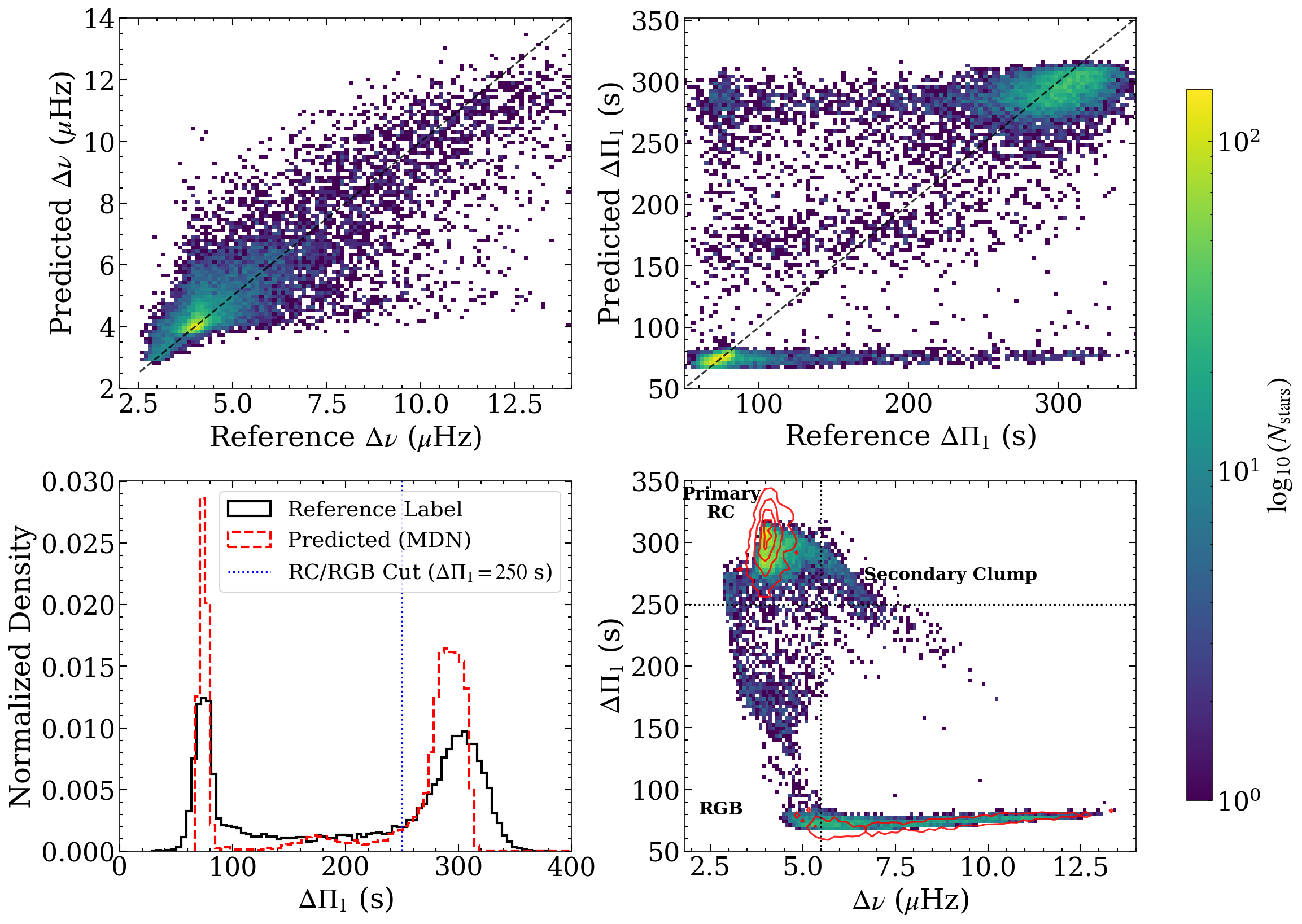}
\caption{MDN prediction performance on the blind test set. Upper left: two-dimensional density distribution of $\Delta\nu_{\rm pred}$ versus $\Delta\nu_{\rm ref}$. Upper right: two-dimensional density distribution of $\Delta\Pi_{1,\rm pred}$ versus $\Delta\Pi_{1,\rm ref}$. The black dashed lines indicate the 1:1 relation. Lower left: histograms of the predicted (red dashed line) and reference-label (black solid line) $\Delta\Pi_1$ values. Lower right: two-dimensional density distribution of predicted (color scale) and reference-label (red contour) $\Delta\Pi_1$ versus $\Delta\nu$. The positions corresponding to $\Delta\Pi_1 = 250$\,s and $\Delta\nu = 5.5\,\mu$Hz are marked.}
\label{fig:mdn}
\end{figure*}

\textcolor{blue}{We first evaluate the performance of the MDN predictions for $\Delta\Pi_1$ and $\Delta\nu$. Fig.~\ref{fig:mdn} presents the results for the independent blind-test set \textcolor{black}{($N = 12{,}452$)}. The predictions recover the distinct RGB-like and RC-like branches without collapsing into the low-density region between them, although the dynamic range within each branch is compressed relative to that of the reference labels. This behavior supports the use of the two-component Gaussian output described above rather than a scalar regression output for the bimodal $\Delta\Pi_1$ distribution. The predicted $\Delta\nu$ values broadly agree with the reference labels over the typical RC range ($3 \lesssim \Delta\nu \lesssim 6\,\mu$Hz). At $\Delta\nu \gtrsim 8\,\mu$Hz, the prediction scatter increases, consistent with the sparse training data in this regime.}

\textcolor{blue}{The distribution of the MDN-predicted $\Delta\Pi_1$ reproduces the bimodal structure of the blind-test reference labels, with an RGB peak near $\Delta\Pi_1 \approx 75$\,s and an RC peak near $\Delta\Pi_1 \approx 300$\,s. The predicted peaks are slightly narrower than the corresponding reference-label distributions. The joint distribution shows three evolutionary regions: RGB stars concentrated at $\Delta\Pi_1 < 100$\,s over a broad range of $\Delta\nu$; primary RC stars concentrated at $\Delta\Pi_1 > 250$\,s and $\Delta\nu \lesssim 5.5\,\mu$Hz; and a high-$\Delta\nu$ extension associated with higher-mass secondary-clump stars.}

\subsection{Primary RC Sample Construction}
\label{sec:rc_samples}

To translate the MDN inference into an all-sky primary RC sample, we impose two criteria, $\textcolor{blue}{\Delta\Pi_{1,\rm pred}} \geq 250$\,s and $\textcolor{blue}{\Delta\nu_{\rm pred}} < 5.5\,\mu$Hz, and use the network-output \textcolor{blue}{\sigdPi} as \textcolor{blue}{the selected-component width and an empirical per-star quality-ranking statistic}. The $\textcolor{blue}{\Delta\Pi_{1,\rm pred}} \geq 250$\,s cut separates core-helium-burning stars from shell-hydrogen-burning RGB stars \citep{bedding2011gravity,mosser2011mixed}.
\begin{figure*}
\centering
\includegraphics[width=0.7\textwidth]{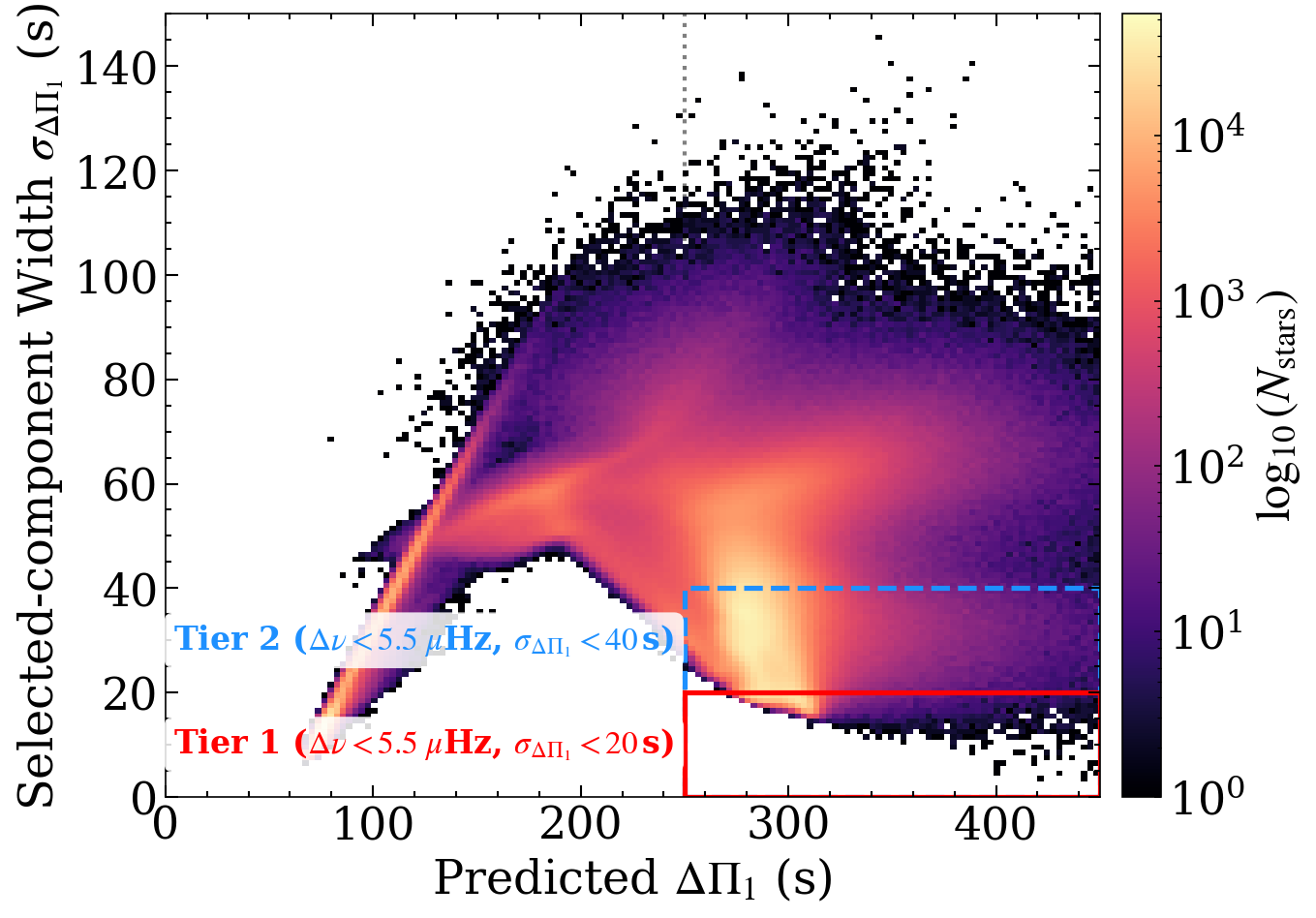}
\caption{\textcolor{blue}{Two-dimensional density distribution of giant candidates with $\Delta\nu_{\rm pred} < 5.5\,\mu$Hz in the $(\Delta\Pi_{1,\rm pred},\,\sigdPi)$ plane, where $\sigdPi$ is the width of the selected Gaussian component. The red solid box marks the Tier~1 high-purity cut ($\Delta\Pi_{1,\rm pred} \geq 250$\,s and $\sigdPi < 20$\,s), and the blue dashed box marks the Tier~2 high-completeness cut ($\Delta\Pi_{1,\rm pred} \geq 250$\,s and $\sigdPi < 40$\,s).}}
\label{fig:sigma}
\end{figure*}

\textcolor{blue}{The primary RC and secondary-clump populations are not separated by a sharp or universal $\Delta\nu$ boundary. Their separation depends on stellar structure and mass, while the limiting mass for non-degenerate helium ignition also depends on chemical composition, including metallicity, and on stellar-model assumptions \citep{girardi1999secondary}. Different seismic classification schemes therefore combine $\Delta\nu$ or $\nu_{\max}$ with mixed-mode information or seismic mass, and classifications can remain ambiguous near the interface \citep{elsworth2019insights}. Because $\Delta\nu$ primarily traces mean density rather than mass directly, we use $\Delta\nu_{\rm pred} < 5.5\,\mu$Hz only as an empirical, sensitivity-tested screening threshold. It preferentially retains lower-mass primary RC stars rather than higher-mass secondary-clump stars and is not interpreted as a universal physical division.}

\textcolor{blue}{Our adopted value is somewhat more permissive than the $\sim 5\,\mu$Hz boundary discussed by \citet{kallinger2012evolutionary} and \citet{mosser2014mixed}. We evaluated the sensitivity to this threshold on the independent blind-test subset of our supervised sample, constructed from the erratum-corrected \citet{ting2018large} reference labels after the catalog cross-matches. Within this blind-test set, tightening the cut to $\Delta\nu_{\rm pred} < 5.0\,\mu$Hz leaves the Tier~1 selection with $\sigma_{\Delta\Pi_1} < 20$\,s unchanged. For the more inclusive Tier~2 selection with $\sigma_{\Delta\Pi_1} < 40$\,s, the stricter cut raises purity only from 84.0\% to 84.5\%, while completeness decreases from 86.0\% to 82.4\%. The stricter cut removes 324 stars from the blind-test Tier~2 selection, of which 241 are reference-label primary RC stars and 83 are reference-label contaminants. We therefore retain $\Delta\nu_{\rm pred} < 5.5\,\mu$Hz because it provides the more favorable purity--completeness trade-off for the high-completeness Tier~2 sample without affecting the Tier~1 blind-test selection.} Fig.~\ref{fig:sigma} shows the two-dimensional density distribution of all $\sim$22 million giant candidates satisfying \textcolor{blue}{$\Delta\nu_{\rm pred}$} $< 5.5\,\mu$Hz in the \textcolor{blue}{$(\Delta\Pi_{1,\rm pred},\,\sigma_{\Delta\Pi_1})$} plane.

We evaluated the purity and completeness of the primary RC sample on the independent blind-test set ($N_{\rm test}=12{,}452$), in which $N_{\rm RC,ref}=6{,}562$ stars are classified as primary RC in the reference-label catalog. Predicted candidates are required to satisfy \textcolor{blue}{$\Delta\Pi_{1,\rm pred}$} $\geq 250$\,s and $\Delta\nu_{\rm pred}<5.5\,\mu$Hz. Applying a strict \textcolor{blue}{selected-component-width threshold} of \textcolor{blue}{$\sigma_{\Delta\Pi_1}$} $<20$\,s results in a sample of 1,928 stars with a purity of 97\% and a completeness of 29\%. Relaxing the threshold to \textcolor{blue}{$\sigma_{\Delta\Pi_1}$} $<30$\,s increases the sample size to 5,534 stars, with purity decreasing to 90\% and completeness rising to 76\%. With \textcolor{blue}{$\sigma_{\Delta\Pi_1}$} $<40$\,s, the sample comprises 6,723 stars and attains a purity of 84\% and a completeness of 86\%. Removing the \textcolor{blue}{$\sigma_{\Delta\Pi_1}$} restriction entirely yields 7,008 stars, corresponding to a purity of 82\% and a completeness of 88\%.

\begin{figure}
\centering
\includegraphics[width=0.5\textwidth]{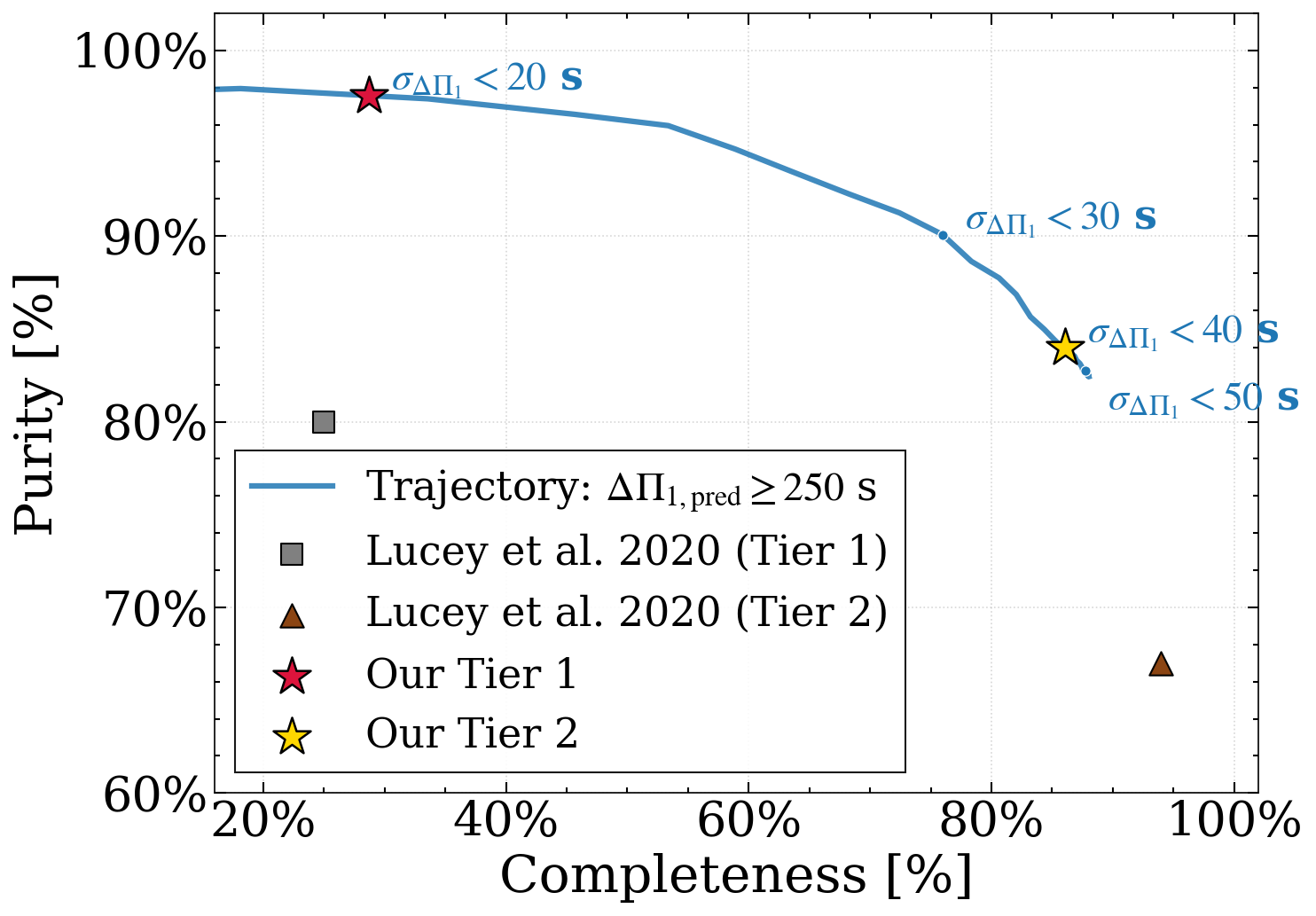}
\caption{\textcolor{blue}{Purity--completeness trade-off for primary RC selection under the common requirements $\Delta\Pi_{1,\rm pred} \geq 250$\,s and $\Delta\nu_{\rm pred} < 5.5\,\mu$Hz. The solid blue curve traces the locus as the $\sigdPi$ threshold is varied; red and yellow stars mark the adopted Tier~1 threshold of 20~s and Tier~2 threshold of 40~s, respectively.} Gray squares and brown triangles show the two sub-samples of \citet{lucey2020inner} based on multi-band photometry.}
\label{fig:purity}
\end{figure}

Based on the above results and considering the purity--completeness trade-off evaluated on the blind test set, we ultimately adopt the strict \textcolor{blue}{$\sigma_{\Delta\Pi_1}$} $<20$\,s cut as the criterion for our high-purity Tier~1 sample, and relax the threshold to \textcolor{blue}{$\sigma_{\Delta\Pi_1}$} $<40$\,s to define our high-completeness Tier~2 sample. Fig.~\ref{fig:purity} places our Tier~1 and Tier~2 on the purity--completeness plane along the locus generated by continuously varying the threshold, and directly compares them with the two sub-samples of \citet{lucey2020inner} based on multi-band photometry. At comparable completeness levels, the \citet{lucey2020inner} high-purity sub-sample (completeness = 25\%) has a purity of 80\%, whereas ours reaches 97\% purity; at the large-sample end, the \citet{lucey2020inner} Tier~2 (completeness = 94\%) declines to $\sim$67\% purity, while our Tier~2 maintains 84\% at 86\% completeness. The \textit{Gaia} XP continuous spectra carry CN molecular-band and continuum-shape information \citep{hawkins2018photospheric} that, relative to broadband photometric SEDs, conveys a more direct evolutionary-branch signal for \textcolor{blue}{$\Delta\Pi_1$} inference.

In comparison with ground-based high- and medium-resolution spectroscopic methods, \citet{ting2018large} \textcolor{blue}{applied their framework to APOGEE ($R \approx 22{,}500$) and LAMOST ($R \approx 1800$) spectra, with the LAMOST sample restricted to ${\rm S/N}_{\rm pix} > 75$, yielding 85,539 RC stars with $\sim97\%$ purity after duplicate correction} \citep{ting2018erratum}; \citet{huang2020mapping} combined LAMOST spectra with \textit{Kepler} asteroseismology to construct approximately 140,000 primary RC stars, typically at $\gtrsim80\%$ purity. Our Tier~1 and Tier~2 samples substantially expand the all-sky high-purity RC sample while maintaining comparable or higher purity, and eliminate the sky selection bias that ground-based spectroscopic surveys suffer from being confined to northern-hemisphere footprints.

\begin{deluxetable*}{lll}
\tablecaption{Description of the columns in the released Tier~1 and Tier~2 primary RC catalogs.\label{tab:catalog}}
\tablewidth{0pt}
\tablehead{
\colhead{Column} & \colhead{Unit} & \colhead{Description}
}
\startdata
\texttt{source\_id}       & \nodata & \textit{Gaia} DR3 source identifier \\
\texttt{ra}               & deg     & Right ascension (Gaia DR3, ICRS) \\
\texttt{dec}              & deg     & Declination (Gaia DR3, ICRS) \\
\texttt{l}                & deg     & Galactic longitude \\
\texttt{b}                & deg     & Galactic latitude \\
\texttt{parallax}         & mas     & \textit{Gaia} DR3 parallax \\
\texttt{parallax\_error}  & mas     & \textit{Gaia} DR3 parallax uncertainty \\
\texttt{teff}             & K       & Effective temperature \citep{zhang2025three} \\
\texttt{logg}             & dex     & Surface gravity \citep{zhang2025three} \\
\texttt{feh}              & dex     & Metallicity [Fe/H] \citep{zhang2025three} \\
\textcolor{blue}{\texttt{DeltaPi1\_pred}}
                         & s
                         & \textcolor{blue}{Location of the maximum-weight $\Delta\Pi_1$ Gaussian component} \\
\textcolor{blue}{\texttt{DeltaPi1\_sigma}}
                         & s
                         & \textcolor{blue}{Width (standard deviation) of that component} \\
\textcolor{blue}{\texttt{DeltaPi1\_pimax}}
                         & \nodata
                         & \textcolor{blue}{Mixing weight $\pi_{k_*}$ of the selected $\Delta\Pi_1$ component} \\
\textcolor{blue}{\texttt{DeltaNu\_pred}}
                         & $\mu$Hz
                         & \textcolor{blue}{Location of the single-Gaussian $\Delta\nu$ output} \\
\textcolor{blue}{\texttt{DeltaNu\_sigma}}
                         & $\mu$Hz
                         & \textcolor{blue}{Width (standard deviation) of that output} \\
\texttt{A\_Ks}            & mag     & $K_s$-band extinction \\
\texttt{d\_phot}          & kpc     & Photometric (standard-candle) distance \\
\texttt{d\_phot\_err}     & kpc     & Photometric distance uncertainty $\sigma_d$ \\
\enddata
\tablecomments{\textcolor{blue}{The full catalogs are publicly available at \url{https://nadc.china-vo.org/res/r101871/}.} A portion is shown here to illustrate the format and content. \textbf{The released files also include \texttt{ruwe}, \texttt{M\_Ks}, and \texttt{d\_phot\_err\_percent} columns not listed above.}}
\end{deluxetable*}

Applying these two sets of thresholds to the all-sky sample of 21,920,705 giant candidates, we obtain Tier~1 with $N=532{,}189$ stars and Tier~2 with $N=6{,}534{,}931$ stars. \textcolor{blue}{Both catalogs are publicly available.} Each catalog entry includes the following columns, summarized in Table~\ref{tab:catalog}: the \textit{Gaia} DR3 source\_id; \textcolor{blue}{the selected-component point estimates $\Delta\Pi_{1,\rm pred}$ and $\Delta\nu_{\rm pred}$, the corresponding component widths $\sigma_{\Delta\Pi_1}$ and $\sigma_{\Delta\nu}$, and the mixing weight $\pi_{k^\ast}$ of the selected $\Delta\Pi_1$ component}; the per-star photometric distance $d_{\rm phot}$ and its error $\sigma_d$; and extinction $A_{K_s}$. The catalogs also include atmospheric parameters (\Teff, \logg, \feh) from the \citet{zhang2025three} catalog and \textit{Gaia} DR3 parallaxes, facilitating user cross-selection.

\subsection{Distance Precision}
\label{sec:dist_results}

The error budget of the RC photometric distance is dominated by three contributions: (i) the intrinsic dispersion of the $\MKs$ calibration, $\sigma_{\MKs} = 0.11$\,mag ($\sim$5\% relative distance error), which is the dominant term; (ii) the fitting-residual dispersion of the intrinsic color, $\sigma_{(J-K_s)_0} = 0.023$\,mag ($\sim$1\%); and (iii) the photometric errors in the 2MASS $J$ and $K_s$ bands ($\sim$2\%). The combined Tier~1 median distance precision is 6\%, consistent with the 5\%--10\% range reported by \citet{huang2020mapping} for LAMOST RCs and with the results of \citet{yu2025stellar}.

\begin{figure*}
\centering
\includegraphics[width=0.85\textwidth]{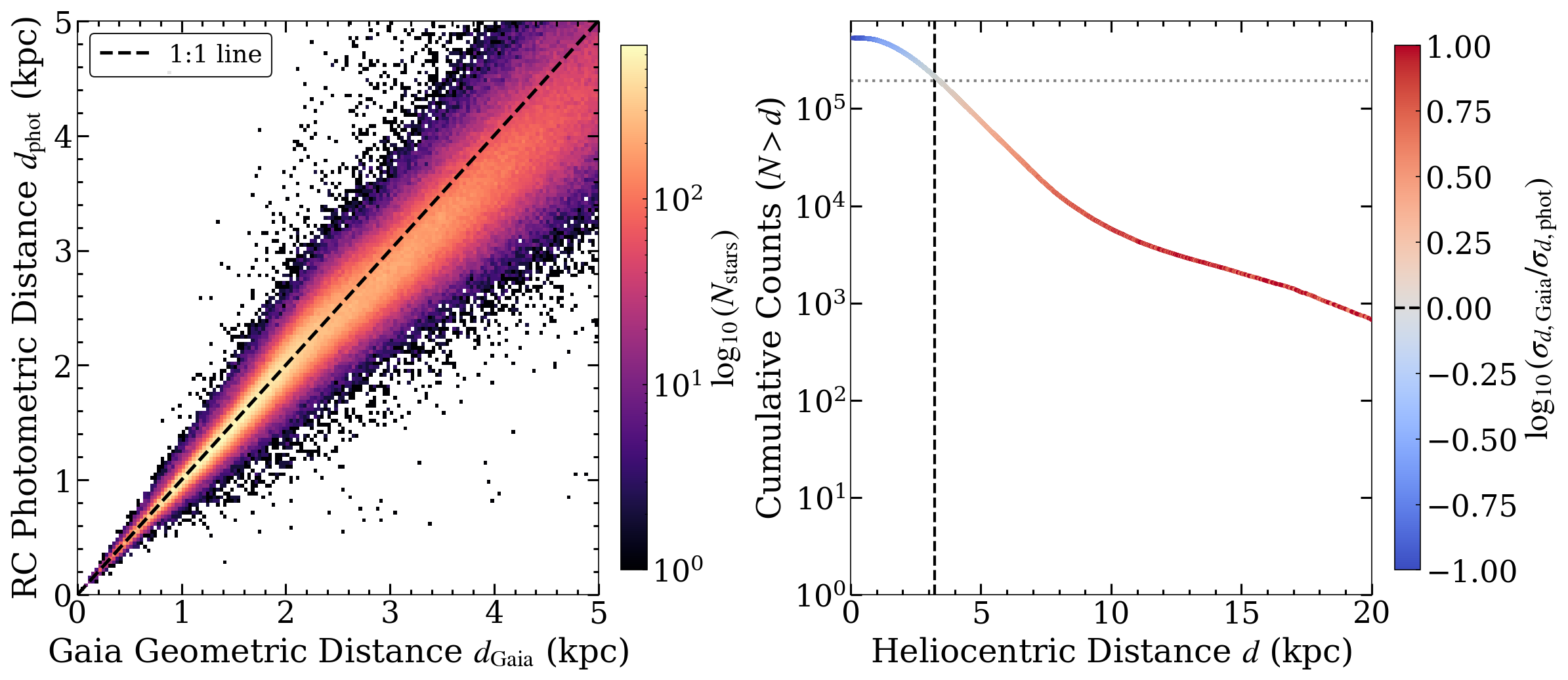}
\includegraphics[width=0.85\textwidth]{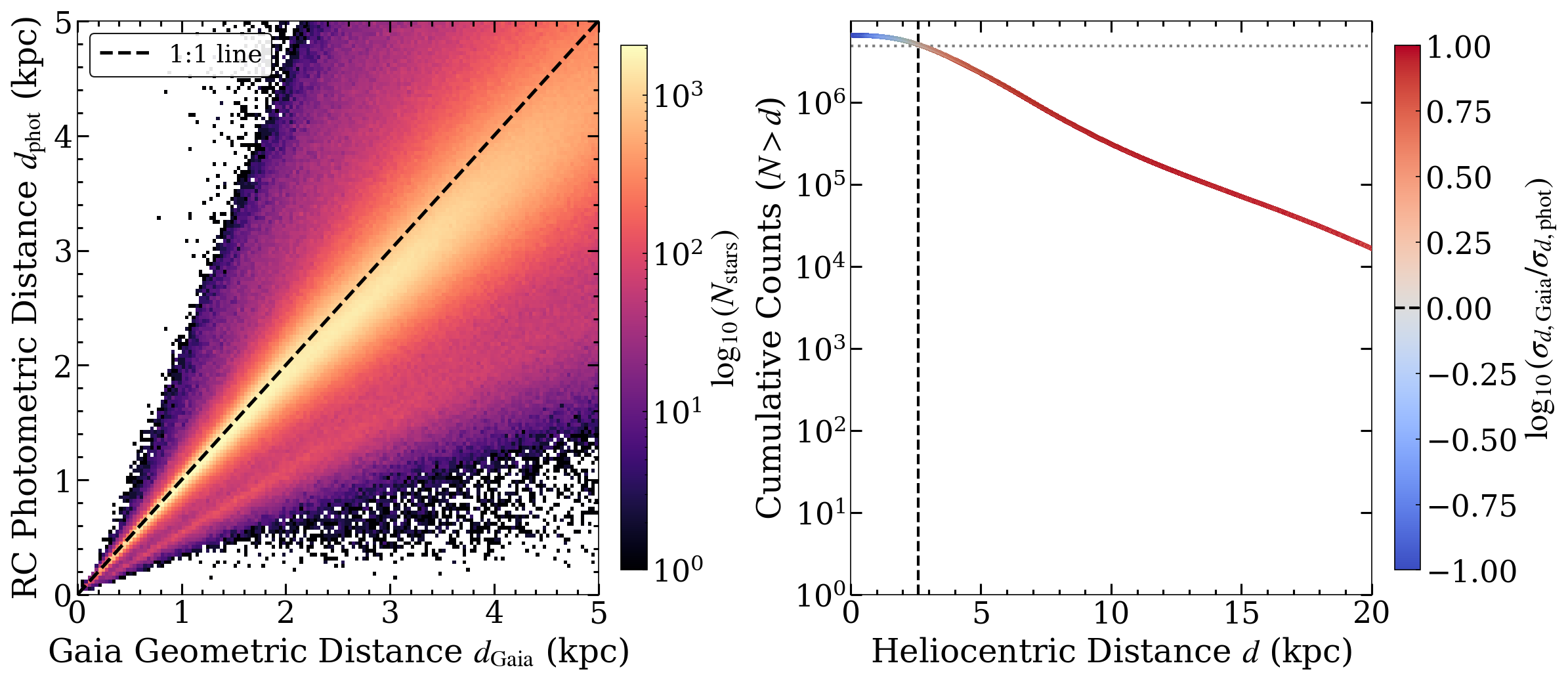}
\caption{Comparison of RC photometric distances with \textit{Gaia} DR3 geometric parallax distances. Left panels: two-dimensional density distributions of $d_{\rm phot}$ versus $d_{\rm Gaia}$ in the near-field regime $d < 5$\,kpc, for the Tier~1 (top) and Tier~2 (bottom) samples. The black dashed lines mark the 1:1 relation. Right panels: cumulative distributions of RC stars with heliocentric distance, \textcolor{blue}{color-coded by $\log_{10}(\sigma_{d,\rm Gaia}/\sigma_{d,\rm phot})$, the logarithm of the ratio of the \textit{Gaia} geometric-distance uncertainty to the RC photometric-distance uncertainty}, for Tier~1 (top) and Tier~2 (bottom). The vertical dashed lines indicate the precision cross-over distances, $d_{\rm cross} = 3.2$\,kpc (Tier~1) and $d_{\rm cross} = 2.6$\,kpc (Tier~2).}
\label{fig:comp2}
\end{figure*}

The photometric distance precision of primary RC stars remains approximately constant with distance, governed by the intrinsic luminosity stability of the standard candle. In contrast, the relative error of \textit{Gaia} geometric parallax distances diverges rapidly with increasing distance. We compare the RC photometric distances $d_{\rm phot}$ with \textit{Gaia} DR3 parallax distances $d_{\rm Gaia}$ ($= 1/\varpi$) on a 1:1 basis in the near-field regime $d < 5$~kpc. We use the inverse parallax for simplicity; we note that a more rigorous treatment would use Bayesian distance estimates that incorporate a Galactic prior, which would reduce the divergence of \textit{Gaia} distance errors at large distances and shift the cross-over distances upward. As shown in the left panels of Fig.~\ref{fig:comp2}, the two distance values agree along the 1:1 line throughout the bulk of the coverage.

\textcolor{blue}{Based on the per-star distance uncertainties}, we define the precision cross-over distance $d_{\rm cross}$ as the distance threshold beyond which the RC photometric distance precision systematically surpasses that of \textit{Gaia} parallaxes ($\sigma_{d_{\rm phot}} < \sigma_{d_{\rm Gaia}}$). The right panels of Fig.~\ref{fig:comp2} display the cumulative distributions for both Tier~1 and Tier~2. For Tier~1, $d_{\rm cross} = 3.2$\,kpc, and 190 thousand primary RC stars lie beyond this distance with photometric precision superior to that of \textit{Gaia} parallaxes. For Tier~2, $d_{\rm cross} = 2.6$\,kpc, and the corresponding count rises to 4.8 million stars. Owing to the high completeness and unbiased all-sky coverage of Tier~2, we deliver a large primary RC sample whose distance precision at intermediate to large Galactocentric distances surpasses that of \textit{Gaia} parallaxes, which is well suited for studies of the outer distant disk. The systematically smaller $d_{\rm cross}$ of Tier~2 relative to Tier~1 reflects two competing effects of the relaxed $\sigdPi < 40$\,s cut. It preferentially admits more distant, low-S/N candidates whose \textit{Gaia} parallax errors are large, shifting the cross-over inward. This dominates over the countervailing effect of increased RGB contamination at all distances, which inflates the photometric distance scatter.

\section{Discussion}
\label{sec:discussion}

This section further demonstrates two direct applications of the sample to Milky Way studies and discusses the current limitations and directions for improvement. 

\subsection{Gaia DR3 Parallax Zero-Point}
\label{sec:zp}

\begin{figure*}
\centering
\includegraphics[width=\textwidth]{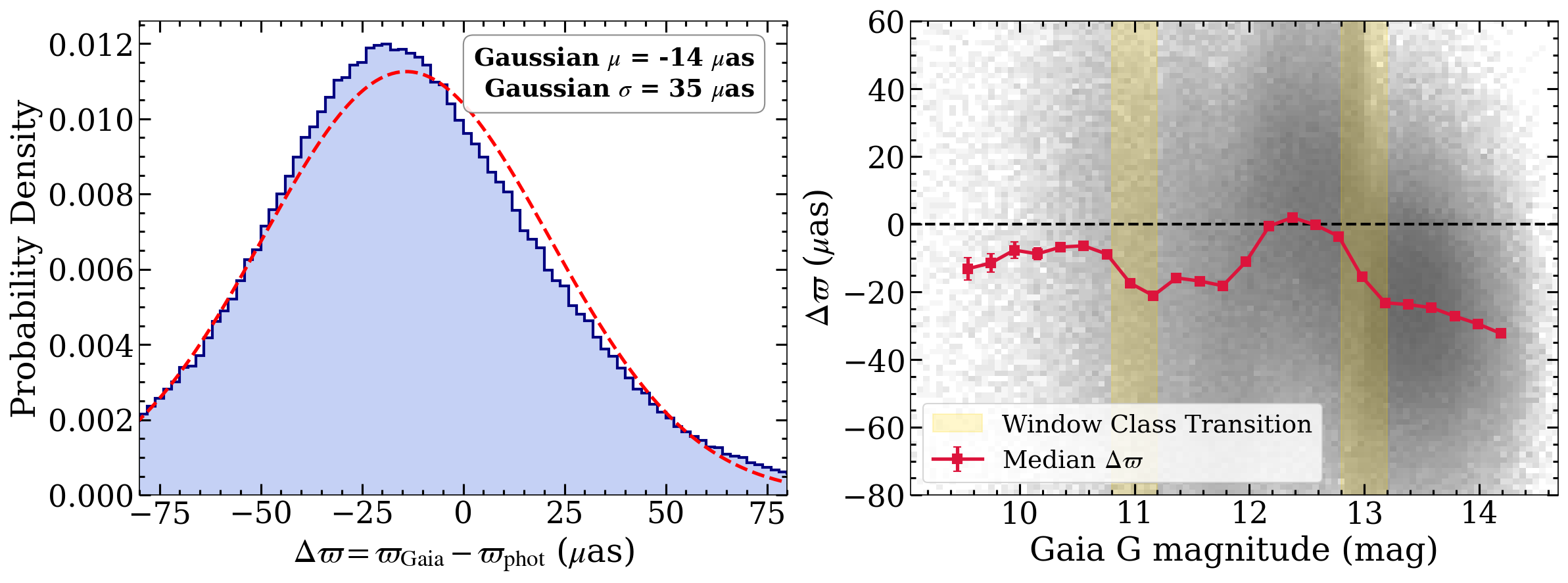}
\caption{\textcolor{blue}{Differential calibration of the effective \textit{Gaia} DR3 parallax offset using the Tier~1 sample. Left: Gaussian fit (red dashed curve) to the parallax residual $\Delta\varpi=\varpi_{\rm Gaia}-\varpi_{\rm phot}$ in the near-field regime $d < 4$\,kpc (blue histogram); the fitted mean is approximately $-14\,\mu$as.} Right: parallax zero-point offset as a function of $G$ magnitude. The grey scales show the stellar density, red squares and curves represent the binned median values, and the horizontal dashed line marks the zero level.}
\label{fig:zeropoint}
\end{figure*}

\textcolor{blue}{Using the photometric parallax $\varpi_{\rm phot}=1/d_{\rm phot}$ from the high-purity Tier~1 sample as an internal photometric reference, we estimate the effective \textit{Gaia} DR3 parallax offset relative to the locally anchored RC distance scale.} To avoid the nonlinear biases that parallax inversion introduces at large distances, we restrict the analysis to the near-field regime $d<4$~kpc and compare $\varpi_{\rm phot}$ against $\varpi_{\rm Gaia}$ on a star-by-star basis for the Tier~1 sample. \textcolor{blue}{As shown in the left panel of Fig.~\ref{fig:zeropoint}, a Gaussian fit to $\Delta\varpi=\varpi_{\rm Gaia}-\varpi_{\rm phot}$ yields a mean residual of approximately $-14\,\mu$as.}

We note that the RC photometric distance scale is ultimately tied to the mean parallax of the local calibrator sample ($\varpi > 1.0$~mas, $d \lesssim 1$~kpc; Section~\ref{sec:dist_method}). The derived zero-point offset therefore represents the effective offset of \textit{Gaia} DR3 parallaxes relative to the mean parallax scale defined by stars within $\sim$1~kpc of the Sun, and should be interpreted as a differential calibration between the local and more distant ($\sim$1--4~kpc) \textit{Gaia} parallax systems rather than an absolute zero-point tied to an external reference frame such as quasars.

\textcolor{blue}{We compare this result with representative studies. \citet{davies2017using} used asteroseismically characterized RC stars in the \textit{Kepler} field and reported a median TGAS-minus-RC parallax residual of approximately $-100\,\mu$as. This value is larger in magnitude than our effective DR3 residual of approximately $-14\,\mu$as, although differences in data release, sample, and calibration convention preclude a direct comparison.} \citet{groenewegen2021parallax} reported a mean zero-point offset of approximately $-21\,\mu$as for \textit{Gaia} EDR3 based on quasars and wide binaries; however, quasar samples are predominantly faint ($G \gtrsim 15$~mag), making it difficult to directly constrain the bright end. \citet{huang2021parallax} used the LAMOST primary RC sample to obtain an offset of approximately $-26\,\mu$as. \textcolor{blue}{The difference of approximately $12\,\mu$as between our offset and the result of \citet{huang2021parallax} may partly reflect the different calibration scales. \citet{huang2021parallax} calibrated RC absolute magnitudes using distances derived from \textit{Gaia} DR2 parallaxes, whereas our $M_{K_s}$ calibration is anchored to \textit{Gaia} DR3 parallaxes within $d \lesssim 1$~kpc. Differences between the DR2 and DR3 parallax systems may therefore contribute. The different sky footprints may also contribute, but the present analysis does not separate these effects.} Our offset should therefore be interpreted as a differential calibration within the \textit{Gaia} DR3 system, complementary to the absolute zero-point derived from quasars.

\textcolor{blue}{Despite the numerical difference in the effective offset,} we reproduce the magnitude dependence of the zero-point reported by \citet{huang2021parallax}. As shown in the right panel of Fig.~\ref{fig:zeropoint}, the calibration exhibits systematic jumps at $G \approx 11$ and 13~mag, coincident with the known instrumental effects associated with \textit{Gaia} CCD gate readout window-class transitions \citep{lindegren2021gaia1,lindegren2021gaia2}. \textcolor{blue}{The recovery of this feature shows that the Tier~1 sample is sensitive to magnitude-dependent astrometric systematics within the adopted differential-calibration framework. Future \textit{Gaia} releases will provide an opportunity to test this calibration against improved astrometry and release-specific systematics.}

\subsection{Spatial Distribution of Tier~2 stars}
\label{sec:spatial}

\begin{figure*}
\centering
\includegraphics[width=\textwidth]{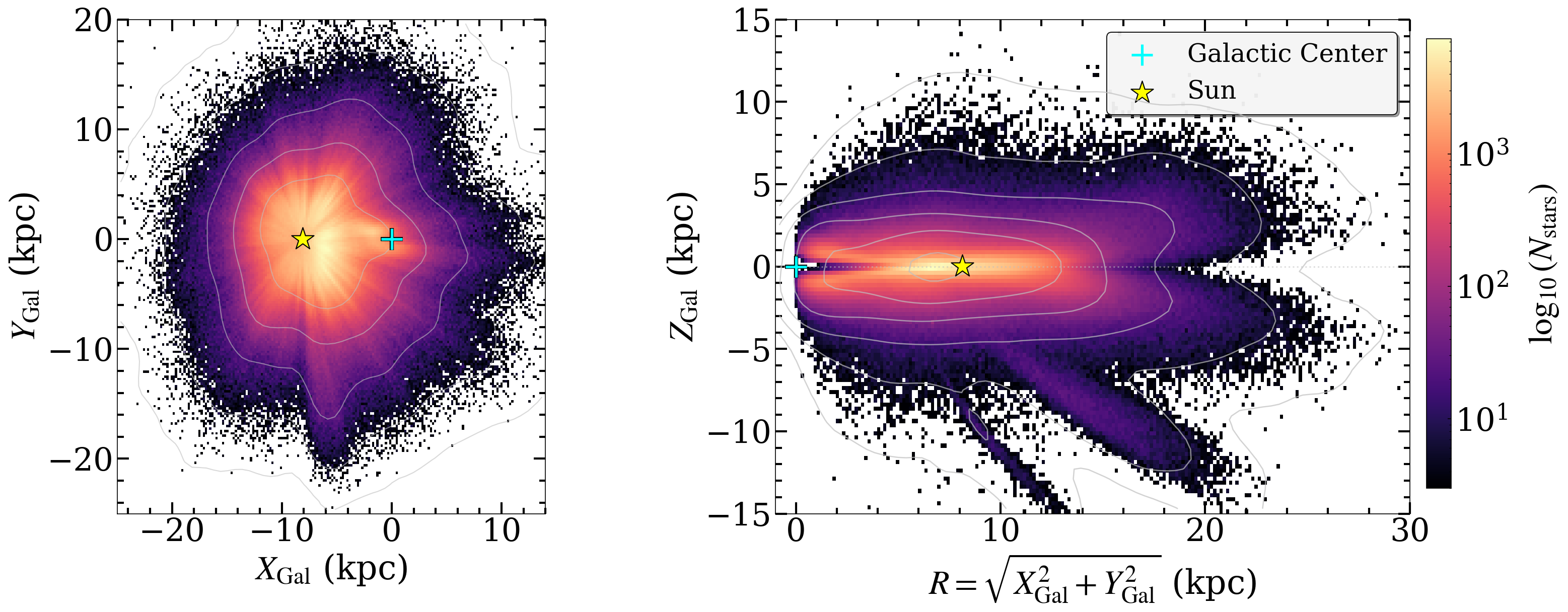}
\caption{Three-dimensional spatial distribution of the Galactic disk traced by the Tier~2 sample. The Sun is located at $R_0 = 8.122$\,kpc and $Z_0 = 20.8$\,pc. Left: face-on view of the Galactic plane ($X_{\rm Gal}$, $Y_{\rm Gal}$); yellow and cyan symbols mark the solar and Galactic center positions, respectively. Right: vertical profile ($R$, $Z$).}
\label{fig:disk}
\end{figure*}

Using the high-completeness Tier~2 sample and its photometric distances, we transform stellar positions into a Galactocentric coordinate system referenced to $R_0 = 8.122$\,kpc \citep{abuter2018detection} and $Z_0 = 20.8$\,pc \citep{bennett2019vertical}. The resulting all-sky three-dimensional spatial distribution is shown in Fig.~\ref{fig:disk}. The left panel presents a face-on view of the Galactic plane ($X_{\rm Gal}$, $Y_{\rm Gal}$). The sample exhibits a continuous, smooth density distribution that traverses the Galactic center direction and extends to the far side of the outer disk. The right panel shows the vertical profile ($R$, $Z$): the stellar density is strongly concentrated near the midplane but extends into a diffuse component at high $|Z|$. In the range $4 \lesssim R \lesssim 12$~kpc, the morphology is qualitatively consistent with a thin-disk/thick-disk two-component geometry. At $R \gtrsim 12$~kpc, the vertical density distribution broadens markedly, consistent with the previously reported disk flaring phenomenon \citep{lopez2002old,yu2021mapping}. Future quantitative density modeling with the Tier~2 sample (e.g., two-component thin-disk plus thick-disk fits to the radial and vertical density profiles) will provide independent constraints on the scale length, the radial variation of the scale height, and the morphological parameters of the flaring (Yu et al., in preparation).

In the lower-right portion of the vertical profile ($10 \lesssim R \lesssim 25$~kpc, $-15 \lesssim Z \lesssim -3$~kpc), two narrow band-like structures extend along the line of sight with geometry converging toward the solar position. Their celestial coordinates concentrate around $(l, b) \approx (280^\circ, -33^\circ)$ and $(303^\circ, -44^\circ)$, coincident with the positions of the Large and Small Magellanic Clouds (LMC/SMC). This residual originates from the sharp S/N drop of \textit{Gaia} XP spectra for LMC/SMC main-body stars ($G \gtrsim 17$~mag). A small fraction of genuine Magellanic Cloud giants, whose XP continua are dominated by photon noise, are misclassified as RC by the MDN. The distance framework then assigns them standard-candle luminosities of $\MKs \approx -1.6$~mag, resulting in their systematic projection to $d \lesssim 25$~kpc and producing a 'finger-of-god' artifact. This residual constitutes $\lesssim 0.5\%$ of the total Tier~2 sample and is strictly confined to the sparse region well outside the disk body, introducing no systematic bias in the statistical inference of outer-disk flaring and thick-disk geometry.

\subsection{Limitations and Future Work}
\label{sec:limitations}

Our method has several limitations, which simultaneously indicate directions for future improvement. First, the asteroseismic training set \citep{ting2018large} is limited by the stellar-population coverage of the \textit{Kepler} field---training samples at the metal-poor end ($\feh \lesssim -1.0$) and the \textcolor{blue}{high-$\Delta\nu$ end} ($\Delta\nu \gtrsim 8\,\mu$Hz) are relatively sparse, leading to systematically larger \textcolor{blue}{prediction scatter} in these parameter domains (see Fig.~\ref{fig:mdn}). \textcolor{blue}{The \textit{PLATO} mission \citep{rauer2014plato} is expected to provide asteroseismic training sets with broader stellar-parameter coverage, helping to alleviate this limitation.} Second, the convex-hull filtering strategy, while effectively preventing extrapolation, unavoidably rejects a small number of \textcolor{blue}{reference-label RC stars} located at the margins of the training-set parameter space; these rejected stars may correspond to populations with distinctive chemical abundances or evolutionary states that merit re-examination under larger training sets.

\textcolor{blue}{Alternative probabilistic architectures provide complementary directions for future work. A Bayesian neural network, or more directly a Bayesian extension of the present MDN through variational weight posteriors or Monte Carlo dropout, could quantify epistemic uncertainty associated with the network parameters, which is not represented by the present deterministic network weights. This would be particularly useful in sparsely sampled regions of the training set. Conditional diffusion models could represent more flexible conditional distributions than the adopted two-component Gaussian mixture, although their iterative inference cost would require careful assessment for an application to approximately 22 million stars. Exact Gaussian-process inference is not practical for our 124,513-object supervised modeling sample with 343-dimensional inputs; scalable inducing-point or deep-kernel approximations, together with a likelihood capable of representing the bimodal $\Delta\Pi_1$ distribution, would be required. We have not performed controlled comparisons among these models and therefore leave their evaluation to future work.}

Third, our distance calibration relies on 2MASS $J$ and $K_s$ photometry---at the faint end ($K_s \gtrsim 13.5$~mag), the 2MASS photometric errors increase rapidly, becoming the bottleneck for distance precision at large heliocentric distances. \textcolor{blue}{Future incorporation of deeper red-optical and near-infrared survey data, such as $y$-band photometry from the Legacy Survey of Space and Time \citep[LSST;][]{ivezic2019lsst} and wide-field near-infrared photometry from the Nancy Grace \textit{Roman} Space Telescope \citep{spergel2015wide}, may extend the effective reach of the RC standard candle.} Finally, our MDN uses the 343-dimensional XP continuous spectrum as the sole input feature, without explicitly incorporating auxiliary information such as metallicity or parallax; a multi-modal input MDN framework is a viable direction for further improving \textcolor{blue}{$\Delta\Pi_1$} inference precision, particularly in the low-S/N regime.

\section{Summary}
\label{sec:conclusion}

\textcolor{blue}{We extended the data-driven asteroseismic inference framework to the all-sky \textit{Gaia} DR3 XP spectra to infer spectroscopic estimates of $\Delta\Pi_1$ and $\Delta\nu$ and to construct a physically motivated primary RC catalog.} We trained an MDN to simultaneously predict \textcolor{blue}{$\Delta\Pi_1$} and $\Delta\nu$ for individual stars, using the 343-dimensional corrected continuous XP spectra of \citet{huang2024comprehensive} as input features, the asteroseismic cross-matched catalog of \citet{ting2018large} as the supervised benchmark, and the all-sky atmospheric parameter catalog of \citet{zhang2025three} as the parent sample. Adopting \textcolor{blue}{$\Delta\Pi_{1,\rm pred}$} $\geq 250$\,s and \textcolor{blue}{$\Delta\nu_{\rm pred}$} $<5.5\,\mu$Hz as selection cuts and the MDN-output \textcolor{blue}{$\sigma_{\Delta\Pi_1}$} as \textcolor{blue}{an empirical per-star quality-ranking criterion}, we constructed two complementary all-sky primary RC catalogs: a high-purity Tier~1 sample of 532,189 stars with a purity of 97\%, and a high-completeness Tier~2 sample of 6.5 million stars with a completeness of 86\%. Using 4,820 local high-precision calibrators, we recalibrated the third-order dependence of \MKs\ on \feh\ and the polynomial relation of \JKS\ with (\Teff, \logg, \feh). After rigorously decoupling the covariance contributions in the distance modulus, the Tier~1 sample achieved a median photometric distance precision of 6\%. The error budget analysis indicates that the intrinsic dispersion of the \MKs\ calibration ($\sim$5\%) is the dominant limiting factor on current precision. Cross-comparison with \textit{Gaia} DR3 geometric parallax distances shows that the RC photometric distances systematically surpass \textit{Gaia} parallax precision beyond $d_{\rm cross}\sim3$~kpc. \textcolor{blue}{Using the Tier~1 sample, we estimated an effective \textit{Gaia} DR3 parallax residual of approximately $-14\,\mu$as relative to the locally anchored RC scale and recovered its magnitude-dependent jumps at $G\approx11$ and 13~mag, which coincide with \textit{Gaia} instrument window-class transitions.} Using the Tier~2 sample, we presented the global three-dimensional stellar density of the Galactic disk from the inner region out to $R\simeq25$~kpc. The distribution qualitatively reveals the stratified morphology of the thin and thick disk and the flaring of the outer disk, in mutual corroboration with recent works.

Looking forward, the framework established here for mapping XP spectra to asteroseismic parameters has broad scope for extension. With \textit{Gaia} DR4 delivering higher-precision BP/RP spectra and improved astrometry, and next-generation asteroseismic missions such as \textit{PLATO} providing training sets with broader parameter coverage, the MDN inference precision and applicability will further improve. Incorporating multi-dimensional input information such as age and chemical abundances, and employing deeper near-infrared photometric data, may push the effective reach of the RC standard candle beyond $\sim 20$\,kpc. The two catalogs released here are publicly available and include MDN-inferred parameters and per-star distances with covariance-decoupled errors. They provide high-quality tracer samples for studies of Galactic structure, chemical evolution, and dynamics.

\section*{Data Availability}

\textcolor{blue}{The two primary RC catalogs presented in this work, the high-purity Tier~1 sample ($N=532{,}189$) and the high-completeness Tier~2 sample ($N=6{,}534{,}931$), are publicly available through the China-VO PaperData repository of the National Astronomical Data Center (NADC; resource no.\ r101871, \url{https://nadc.china-vo.org/res/r101871/}). All input data are publicly available: the \textit{Gaia} DR3 XP spectra from the \textit{Gaia} archive, the corrected XP spectra from \citet{huang2024comprehensive}, the atmospheric parameter catalog from \citet{zhang2025three}, and the asteroseismic training labels from the erratum-corrected catalog of \citet{ting2018large} \citep{ting2018erratum}.}

\section*{Acknowledgments}
\textcolor{blue}{We thank the referee for constructive comments that improved the clarity of the manuscript and Dr.~Jie Yu for helpful discussions.} This work is supported by the National Natural Science Foundation of China 12322304, National Natural Science Foundation of Yunnan Province 202301AV070002, and the Xingdian talent support programme of Yunnan Province. We acknowledge the science research grants from the China Manned Space Project with no.~CMS-CSST-2025-A11.

This work has made use of data from the European Space Agency (ESA)
mission \textit{Gaia} (\url{https://www.cosmos.esa.int/gaia}), processed by the \textit{Gaia} Data Processing and Analysis Consortium (DPAC, \url{https://www.cosmos.esa.int/web/gaia/dpac/consortium}). Funding for the DPAC has been provided by national institutions, in particular the institutions participating in the \textit{Gaia} Multilateral Agreement.

This publication makes use of data products from the Two Micron All Sky
Survey, which is a joint project of the University of Massachusetts and the Infrared Processing and Analysis Center/California Institute of Technology, funded by the National Aeronautics and Space Administration and the National Science Foundation.

Guoshoujing Telescope (the Large Sky Area Multi-Object Fiber Spectroscopic Telescope, LAMOST) is a National Major Scientific Project built by the Chinese Academy of Sciences. Funding for the project has been provided by the National Development and Reform Commission. LAMOST is operated and managed by the National Astronomical Observatories, Chinese Academy of Sciences.



\bibliographystyle{aasjournal}
\bibliography{refs}

\end{document}